\documentclass[twocolumn,epjc3]{svjour3}

\smartqed  

\usepackage[pdftex]{graphicx}
\usepackage{mathptmx}      
\usepackage[colorlinks,citecolor=blue,urlcolor=blue,linkcolor=blue]{hyperref}
\usepackage{amsmath,amssymb}
\usepackage{array,booktabs}
\usepackage{rotating}
\usepackage{tablefootnote,threeparttable}
\usepackage{breqn}
\usepackage{cite}
\usepackage{url}
\usepackage{multirow}
\RequirePackage{latexsym}
\journalname{Eur. Phys. J. C}
\begin{document}

\title{Some new Wyman-Adler type static relativistic charged anisotropic fluid spheres compatible to \emph{self-bound} stellar modeling
}

\titlerunning{Relativistic charged anisotropic fluid spheres}        

\author{Mohammad Hassan Murad\thanksref{e1,addr1}
        \and
        Saba Fatema\thanksref{e2,addr2} 
}

\thankstext{e1}{e-mail: mhmurad@bracu.ac.bd}
\thankstext{e2}{e-mail: saba@daffodilvarsity.edu.bd}

\authorrunning{M.H. Murad and S. Fatema} 

\institute{Department of Mathematics and Natural Sciences, BRAC University, Dhaka, 1212, Bangladesh \label{addr1}
           \and
           Department of Natural Sciences, Daffodil International University, Dhaka, 1207, Bangladesh \label{addr2}
}

\date{Received: date / Accepted: date}

\maketitle

\begin{abstract}
In this work some families of relativistic anisotropic charged fluid spheres have been obtained by solving Einstein-Maxwell field equations with the preferred form of one of the metric potentials, a suitable forms of electric charge distribution and pressure anisotropy functions. The resulting equation of state (EOS) of the matter distribution has been obtained. Physical analysis shows that the relativistic stellar structure for matter distribution obtained in this work may reasonably model an electrically charged compact star whose energy density associated with the electric fields is on the same order of magnitude as the energy density of fluid matter itself (e.g. electrically charged \emph{bare} strange stars). These models permit a simple method of systematically fixing bounds on the maximum possible mass of cold compact electrically charged self-bound stars. It has been demonstrated numerically that the maximum compactness and mass increase in the presence of electric field and anisotropic pressures.\par
Based on the analytic model developed in this present work, the values of the relevant physical quantities have been calculated by assuming the estimated masses and radii of some well known potential strange star candidates like PSR J1614-2230, PSR J1903+327, Vela X-1, and 4U 1820-30.

\end{abstract}

\section{Introduction}\label{Sec1}

The subject of modeling relativistic compact stellar objects through the analytical solution of Einstein's gravitational field equations has a long history and still the interest remains as one of the key issue to the present researchers. Since the work of Schwarzschild \cite{Schwarzschild03}, Tolman \cite{Tolman39} and Oppenheimer and Volkoff \cite{Oppenheimer-Volkoff39} the determination of maximum mass of very compact astrophysical objects has been a key issue in relativistic astrophysics. Such findings are important in because analytical solutions enable the distribution of matter in the interior of stellar object to be modeled in terms of simple algebraic relations. \par

The central energy density of compact stellar object could be of the order of $10^{15}$g cm$^{-3}$, several times higher the normal nuclear matter density and due to the absence of reliable information about behavior of matter at such ultrahigh density, insight into the structure can be obtained by reference to applicable analytic solutions to the equation of relativistic stellar structure \cite{Lattimer-Prakash01}. The known analytic solutions of Einstein's gravitational field equations fall into two classes. The first class that describes ``normal'' matter neutron stars for which density vanishes at the surface where the pressure vanishes. The Tolman VII solution with vanishing surface energy density falls into this class and hence is useful approximation to realistic neutron star models. And the class that describes stars for which density is finite, about 2-3 times the normal nuclear matter saturation density \cite{Postnikovetal.10}, at the surface where the pressure vanishes includes Tolman IV solution and the solutions discussed in \cite{Wyman49,Heintzmann69,Adler74,Adams-Cohen75,Kuchowicz75,Vaidya-Tikekar82,Korkina81,Durgapal82,
Orlyansky97,Durgapal-Bannerji83,Durgapaletal.84,Durgapal-Fuloria85,Pant-Pant93,Pant94,Pant96,Maurya-Gupta12a}. This type of solutions is useful approximation to realistic models of ``self-bound'' strange quark star \cite{Lattimer04}. The best-known example of self-bound stars results from the Bodmer-Witten hypothesis also known as the strange quark matter hypothesis asserts that strange quark matter (SQM for short) is the ultimate ground state of matter. Still the fundamental significance of this hypothesis remains as a serious possibility in physics and astrophysics \cite{Weber99,Weber05,Glendenning00,Haensel03,Haenseletal.07}.\par

An important distinction between quark stars and conventional neutron stars is that the quark stars are self-bound by the strong interaction, gravity just make them massive, whereas neutron stars are bound by gravity. This allows a quark star to rotate faster than would be possible for a neutron star. A quark star can also be \emph{bare}. The surfaces of a bare strange star and that of normal matter neutron star have striking differences. The very properties of the quark surface, e.g., strong bounding of particles, abrupt density change from $4\times10^{14}$g cm$^{-3}$ to $\sim0$ in $\sim1$ fm. \par

In a very recent past a polytropic quark star model has been suggested \cite{Lai-Xu09,Thirukkanesh-Ragel12} in order to establish a general framework in which theoretical quark star models could be tested by observations. The key difference between polytropic quark stars and the polytropic model studied previously for normal (i.e. non-quarkian) stars is that the quark star models with non-vanishing density at the stellar surface may not be avoidable due to the strong interaction between quarks which is relevant to the effect of color confinement. As discussed in \cite{Lai-Xu09} the polytropic equations of state are stiffer than the conventional realistic models (e.g. the MIT bag model) for quark matter, and pulsar-like stars calculated with a polytropic equation of state could then have high maximum masses $>2M_\odot$. In this framework of polytropic model a very low massive quark star can also be, and be still gravitationally stable even if the polytropic index, $n>3$. \par
Apart from the constituents of these types of compact stars, the most fascinating distinction between a strange star and a normal neutron star is the surface electric fields associated with it. Bare strange stars possess ultra-strong electric fields on their surfaces, which, for ordinary strange matter, is around $10^{18}$ V/cm and $10^{20}$ V/cm for color superconducting strange matter \cite{Usov04,Usovetal.05,Negreirosetal.10}. The influence of energy densities of ultra-high electric fields on the bulk properties of compact stars was explored in \cite{Rayetal.03,Malheiroetal.04,Weberetal.07,Weberetal.09,Weberetal.10,Negreirosetal.09}. It also has been shown that electric fields of this magnitude, generated by charge distributions located near the surfaces of strange quark stars, increase the stellar mass by up to 30\% depending on the strength of the electric field. In contrast to the strange star the surface electric field in the case of neutron star is absent \cite{Weberetal.12}. These features may allow one to observationally distinguish quark stars from neutron stars. \par
The principal motivation of this work is to develop some new analytical relativistic stellar models by obtaining closed-form solutions of Einstein-Maxwell field equations. In order to obtain a realistic charged stellar model one can start with an explicit EOS and suitable form of electric charge distribution and then integrating the equation of hydrostatic equilibrium, also known charged generalization of Tolman-Oppenheimer-Volkoff (TOV) equation \cite{Bekenstein71} which is obtained by requiring the conservation of mass-energy, as that determines the global structure of electrically charged stars. The integration starts at the center of the star with a prescribed central pressure and ends where the pressure decreases to zero, indicates the surface of the star. Some recent studies include \cite{Feliceetal.95,Anninos-Rothman01,Siffertetal.07,Negreiros-Malheiro07}. Such input equations of state do not normally allow for closed-form solutions. \par

In the second approach one can have insight into such structures by solving the Einstein-Maxwell equations which represent an under-determined system of nonlinear ordinary differential equations of the second order. Due to the high nonlinearity it is difficult to obtain exact solutions to this system. For the special case of a static isotropic perfect fluid, the system of field equations can be reduced to a set of four coupled ordinary differential equations in five unknowns and arrive to exact solutions by making an \emph{ad hoc} assumption for one of the metric functions or for the energy density. The EOS can then be extracted from the resulting metric. The first exact solutions of field equations, in this approach, known to have astrophysical significance, may have been discovered by Tolman \cite{Tolman39}. Out of the different types of exact solutions obtained by Tolman, model V and VI are not considered physically viable, as they correspond to singular solutions (infinite values of central density and pressure). Except these models, all other solutions are known as regular solutions (finite and positive pressure and density at the origin). Models IV and VII are found physically viable in the study of compact astrophysical stellar objects. Out of numerous works done following Tolman's approach some include \cite{Nduka76,Nduka78,Mak-Fung95,Maketal.96,Pateletal.97,Harko-Mak00,Sharmaetal.01,Sharmaetal.06,Gupta-Kumar05a,Gupta-Kumar05b,
Gupta-Kumar05c,Maharaj-Komathiraj07,Thirukkanesh-Maharaj09,Gupta-Kumar11a,Gupta-Kumar11b,Hansraj-Maharaj06, Pant11a,Pant11b,Pant-Tewari11,Pantetal.11a,Pantetal.11b,Pant12,Pant-Maurya12,Pant-Faruqi12,Pant-Negi12,Maurya-Gupta12b,Mauryaetal.12,Bijalwan12,Thirukkanesh-Ragel12,Takisa-Maharaj12,Takisa-Maharaj13a,Takisa-Maharaj13b,Hansrajetal.14}. As might be expected with Tolman's method, unphysical pressure-density configurations are found more frequently than physical ones. \par

In recent years, however, several authors followed an alternative approach to present analytical stellar models of electrically neutral/charged compact strange stars within the framework of linear equation of state (EOS) based on MIT bag model together with a particular choice of metric potentials/mass function \cite{Mak-Harko04,Sharma-Maharaj07,Esculpi-Aloma10,Komathiraj-Maharaj11,Takisa-Maharaj12,
Maharaj-Takisa12,Rahamanetal.12,Kalametal.13,Thirukkanesh-Ragel13b}. Some works also studied the viability of nonlinear EOS based on suitable geometry for the description in the interior 3-spaces of such compact star \cite{Vaidya-Tikekar82,Tikekar90}. This approach leads to physically viable and easily tractable models of superdense stars in equilibrium. Tikekar and Jotania \cite{Tikekar-Jotania05,Tikekar-Jotania07}, Jotania and Tikekar \cite{Jotania-Tikekar06} showed that the ansatz suggested by Tikekar and Thomas \cite{Tikekar-Thomas98} has these features and the general three-parameter solution based on it also leads to physically plausible relativistic models of strange stars. Several aspects of physical relevance and the maximum mass of class of compact star models, based on Vaidya-Tikekar ansatz, for the both isotropic and anisotropic pressures have been investigated in \cite{Tikekar-Thomas99,Sharmaetal.06,Debetal.12,Sharma-Ratanpal13}. \par

Out of the 127 known analytical solutions to Einstein's equations, compiled in \cite{Delgaty-Lake98}, only a few satisfy elementary tests of physical relevance and, hence, are viable in the description of relativistic compact stellar objects. For strange quark stars, the energy density does not vanish at the surface. Known applicable analytic solutions include \cite{Lattimer-Prakash01,Postnikovetal.10,Lattimer04}

\begin{itemize}
\item Schwarzschild interior solution or the incompressible fluid solution (constant density solution).
\item Generalized Tolman IV solution.
\item Matese and Whitman I.
\end{itemize}

In contrast, so far the literature known to present authors, the charged analogues of Tolman's models (V-VI) obtained in \cite{Pant-Sah79,Patino-Rago89,Singhetal.95,Ray-Das02,Ray-Das04,Ray-Das07,Rayetal.07} are not physically viable in the description of compact astrophysical objects as the infinite values of central density and pressure. Though the Schwarzschild constant density solution is physically unrealistic, the charged analogues, obtained in \cite{Guptaetal.12,Gupta-Kumar05c,Bijalwan-Gupta08,Bijalwan11}, and the charged analogue of Matese and Whitman solution obtained in \cite{Maurya-Gupta12c} may be relevant in the description of self-bound electrically charged strange quark stars. Charged analogues of Tolman IV and VII models \cite{Gupta-Maurya11,Maurya-Gupta11c,Kiess12}), as the neutral ones, exhibit the physical features required for the construction of physically realizable relativistic compact stellar structure. The charged analogues of Vaidya-Tikekar models have been derived in \cite{Patel-Koppar87,Kopparetal.91}. Astrophysical consequences of the charged analogues of Vaidya-Tikekar solutions in modeling electrically charged compact star have been discussed in \cite{Gupta-Kumar11a,Patel-Pandya86,Komathiraj-Maharaj07,Bijalwan-Gupta11,Chattopadhyayetal.12}.\par

It was shown by Bonnor \cite{Bonnor60,Bonnor65} that a spherical body can remain in equilibrium under its own gravitation and electric repulsion if the matter present in the sphere carries certain modest electric charge density. The problem of the stability of a homogeneous distribution of matter containing a net surface charge was considered by Stettner \cite{Stettner73}. Stettner showed that a fluid sphere of uniform density with a modest surface charge is more stable than the same system without charge. The electric charge weakens gravity to the extent of turning it into a repulsive field, as happens in the vicinity of a Reissner-Nordstr\"{o}m singularity. Thus the gravitational collapse of spherical matter distribution to a point singularity may be avoided if the matter acquires large amounts of electric charge during an accretion process onto a compact object. The gravitational attraction may then be balanced by electrostatic repulsion due to the same electric charge and by the pressure gradient \cite{Bekenstein71,Ghezzi05}. And hence the study of the gravitational behavior of compact charged stellar object has raised the possibility of modeling such compact astrophysical objects in terms of simple algebraic relations between the matter pressure and its energy density. \par

Of course no astrophysical object is entirely composed of perfect fluid. The theoretical investigations of Ruderman \cite{Ruderman72} about more realistic stellar models show that the nuclear matter may be locally anisotropic at least in certain very high density ranges $(\rho> 10^{15}$ g cm$^{-3})$, where the nuclear interactions in the stellar matter must be treated relativistically. According to these views, in such massive stellar objects the radial pressure may not be equal to the tangential pressure. Since the pioneering work of Bowers and Liang \cite{Bowers-Liang74}, there has been an extensive literature devoted to the study of anisotropic spherically symmetric static general relativistic con­figurations (See \cite{Cosenzaetal.81,Bayin82,Herrera-Leon85,Leon87a,Leon87b,Bondi92,Herrera-Santos97,Herreraetal.01,Dev-Gleiser02,Dev-Gleiser03,Gleiser-Dev04,
Mak-Harko02a,Mak-Harko02b,Maketal.02,Mak-Harko03,Chaisi-Maharaj06a,Chaisi-Maharaj06b,Sharma-Maharaj07,Herreraetal.08,
Viaggiu09,Maharaj-Takisa12,Maurya-Gupta12d,Maurya-Gupta13,Takisa-Maharaj13a,Takisa-Maharaj13b,Maurya-Gupta14,
Maharaj-Sunzu-Ray14,Sunzu-Maharaj-Ray14} and the references therein).\par

Following the approach of Durgapal \cite{Durgapal82}, Maurya and Gupta \cite{Maurya-Gupta11b,Maurya-Gupta11c} some new analytical relativistic stellar models have been developed by obtaining closed-form solutions of Einstein-Maxwell field equations. Our analysis depends on several mathematical key assumptions. First, we choose a particular functional form for one of the metric potentials. The form chosen ensures that the metric function is nonsingular, continuous, and well behaved in the interior of the star. On physical basis this is one of the desirable features for any well-behaved model. Further, we assume particular forms of electric charge distribution and pressure anisotropy. The maximum allowable mass and corresponding values of physical quantities have been determined. The solutions obtained in this work are expected to provide simplified but easy to mathematically analyzed charged stellar models with nonzero super-high surface density which could reasonably model the stellar core of an electrically charged strange quark star by satisfying applicable physical boundary conditions. \par

The presentation of this work is as follows. The next section, Sect. \ref{Sec2}, is devoted for the solution of Einstein-Maxwell field equations of perfect fluid and derives the pressure and density relation. In Sect. \ref{Sec3} we present the elementary criteria to be satisfied the obtained solution as to present a realistic stellar model. Sect. \ref{Sec4} develops the important ratios by matching the obtained metric components with the space-time exterior to the charged object which is described by the unique Reissner-Nordstr\"{o}m metric. Physical analysis has been made on the obtained models in Sect. \ref{Sec5}. It has been demonstrated numerically that the maximum compactness, redshift and mass increase in the presence of electric field and anisotropic pressures; which are in agreement with some other work\cite{Karmakaretal.07}. In Sect. \ref{Sec6} some explicit numerical models of relativistic anisotropic stars, with a possible astrophysical relevance, are also presented and we also apply our ``toy" model to some well known potential strange star candidates to calculate various physical quantities by assuming the estimated masses and predicted radii. And finally Section \ref{Sec7} discusses and concludes the work.\par

\numberwithin{equation}{section}
\section{Interior Solutions of Einstein-Maxwell Field Equations}\label{Sec2}
\numberwithin{equation}{subsection}
\subsection{Field equations}\label{Sec2.1}

In this work we intend to study a static, spherically symmetric matter distribution whose interior metric is given in Schwarzschild coordinates\cite{Tolman39,Oppenheimer-Volkoff39} $x^\mu=\left(t,r,\theta,\varphi\right)$\footnote[1]{Throughout the work we will use $c=G=1$, except in tables and figures.},
\begin{equation}\label{2.1.1}
ds^2=e^{\nu(r)}dt^2-e^{\lambda(r)}dr^2-r^2\left(d\theta^2+\sin^2\theta d\varphi^2\right)
\end{equation}\noindent
The functions $\nu$ and $\lambda$ satisfy the Einstein-Maxwell field equations,
\begin{equation}\label{2.1.2}
G_\nu^\mu=R_\nu^\mu-\frac{1}{2}\delta_\nu^\mu=\kappa\left(T_\nu^\mu+E_\nu^\mu\right)
\end{equation}\noindent
where $\kappa=8\pi$ is Einstein's constant. The matter within the star is assumed to be locally anisotropic fluid in nature and consequently $T_\nu^\mu$  and $E_\nu^\mu$ are the energy-momentum tensor of fluid distribution and electromagnetic field defined by, \cite{Dionysiou82},

\begin{eqnarray*}
T_\nu^\mu &=&\left(P_t+\mu c^2\right)v^\mu v_\nu-P_t\delta^\mu_\nu+(P_r-P_t)\chi^\mu\chi_\nu\\
E_\nu^\mu &=&\frac{1}{4\pi}\left(-F^{\mu m} F_{\nu m}+\frac{1}{4}\delta_\nu^\mu F^{m n} F_{m n}\right)
\end{eqnarray*}
where $\rho,\,P_r,\,P_t,\, v^\mu$, denote energy density, radial pressure, tangential pressure of the fluid distribution respectively. $v^\mu$ and $F_{\mu\nu}$  denote the velocity vector and anti-symmetric electromagnetic field strength tensor defined by,
\begin{equation}\label{2.1.3}
F_{\mu\nu}=\frac{\partial A_\nu}{\partial A_\mu}-\frac{\partial A_\mu}{\partial A_\nu}
\end{equation}\noindent
which satisfies Maxwell equations,
\begin{subequations}
\begin{eqnarray}
F^{\mu\nu}_{;\nu}=\frac{1}{\sqrt{-g}}\frac{\partial}{\partial x^\nu}\left(\sqrt{-g}F^{\mu\nu}\right)&=&-4\pi j^\mu\label{2.1.4a}\\
F_{\mu\nu;\lambda}+F_{\nu\lambda;\mu}+F_{\lambda\mu;\nu}&=&0\label{2.1.4b}
\end{eqnarray}
\end{subequations}\noindent
where $g$ is the determinant of quantities $g_{\mu\nu}$ in eq. \eqref{2.1.1}, defined by,
\begin{equation*}
g=
\begin{pmatrix}
  e^\nu & 0 & 0 & 0 \\
  0 & -e^\lambda & 0 & 0 \\
  0 & 0 & -r^2 & 0 \\
  0 & 0 & 0 & -r^2\sin^2\theta \\
\end{pmatrix}
=-e^{\nu+\lambda}r^4\sin^2\theta
\end{equation*}\noindent
where, $A_\nu=(\varphi(r),0,0,0)$ is four potential and $j^\mu$ is the four current vector defined by
\[j^\mu=\frac{\rho_{\mathrm{ch}}}{\sqrt{g_{00}}}\frac{dx^\mu}{dx^0}\]\noindent
where $\rho_{\mathrm{ch}}$ denotes the proper charge density.\\

For static matter distribution the only non-zero component of the four-current is $j^0$. Because of spherical symmetry, the four-current component is only a function of radial distance, $r$. The only nonvanishing components of electromagnetic field tensor are $F^{01}$ and $F^{10}$, related by $F^{01}=-F^{10}$, which describe the radial component of the electric field. From eq. \eqref{2.1.4a} one obtains the following expression for the electric field:
\[F^{01}=-e^{-\frac{\nu+\lambda}{2}}\frac{q(r)}{r^2}\]\noindent
where $q(r)$ represents the total charge contained within the sphere of radius $r$ defined by,
\begin{equation}\label{2.1.5}
q(r)=4\pi\int_0^r{e^{\frac{\lambda}{2}}\rho_{\mathrm{ch}}u^2}du
\end{equation}\noindent
Equation \eqref{2.1.5} can be treated as the relativistic version of Gauss's law.

For the metric \eqref{2.1.1}, the Einstein-Maxwell field equations may be expressed as the following system of ordinary differential equations \cite{Dionysiou82},

\begin{eqnarray}
\frac{\nu^'}{r}e^{-\lambda}-\frac{\left(1-e^{-\lambda}\right)}{r^2} &=&\kappa P_r-\frac{q^2}{r^4} \label{2.1.6} \\
\left(\frac{\nu^{''}}{2}-\frac{\nu^'\lambda^'}{4}+\frac{\nu^{'2}}{4}+\frac{\nu^'-
\lambda^'}{2r}\right)e^{-\lambda}  &=& \kappa P_t+\frac{q^2}{r^4}\label{2.1.7} \\
\frac{\lambda'}{r}e^{-\lambda}+\frac{\left(1-e^{-\lambda}\right)}{r^2} &=&\kappa\rho+\frac{q^2}{r^4} \label{2.1.8}
\end{eqnarray}\noindent
where prime $\left('\right)$ denotes the $r$-derivative.\par
In analogy to the electrically uncharged case, it is usually introduced a quantity $m(r)$ by the following expression,
\begin{equation}\label{2.1.9}
e^{-\lambda}=1-\frac{2m(r)}{r}+\frac{q^2}{r^2}
\end{equation}
If $R$ represents the radius of the fluid distribution then it can be showed that $m$ is constant $m(r=R)=M$ outside the fluid distribution where $M$ is the gravitational mass. Thus the function $m(r)$ represents the gravitational mass of the matter contained in a sphere of radius $r$. Using eqs. \eqref{2.1.8}, \eqref{2.1.6}, and \eqref{2.1.7} respectively, one can arrive at,
\begin{equation}\label{2.1.10}
m(r)=\frac{\kappa}{2}\int\rho r^2 dr+\frac{q^2}{2r}+\frac{1}{2}\int \frac{q^2}{r^2}dr
\end{equation}
\begin{equation}\label{2.1.11}
\nu'=\frac{(\kappa rP_r+\frac{2m}{r^2}-\frac{2q^2}{r^3})}{(1-\frac{2m}{r}+\frac{q^2}{r^2})}
\end{equation}
\begin{eqnarray}\label{2.1.12}
\frac{dP_r}{dr}=-\frac{(P_r+\rho)}{2}\nu'+\frac{q}{4\pi r^4}\frac{dq}{dr}+\frac{2\Delta}{r}
\end{eqnarray}\noindent
Finally combining \eqref{2.1.11} and \eqref{2.1.12}, we get,
\begin{eqnarray}\label{2.1.13}
\frac{dP_r}{dr}&=&-\frac{(P_r+\rho)}{2}\frac{(\kappa rP_r+\frac{2m}{r^2}-\frac{2q^2}{r^3})}{(1-\frac{2m}{r}+\frac{q^2}{r^2})}\nonumber\\
&&+\frac{q}{4\pi r^4}\frac{dq}{dr}+\frac{2\Delta}{r}
\end{eqnarray}\noindent
which is the charged generalization of Tolman-Oppenheimer-Volkoff (TOV) equation of hydrostatic equilibrium for anisotropic stellar configuration \cite{Leon87b}.
In eq. \eqref{2.1.13} the additional term, $2\Delta/r$, representing ``force'' which is due to the anisotropic nature of the fluid. This force is directed outward when $P_t>P_r\,(\Delta>0)$ and inward when $P_t<P_r\,(\Delta<0)$. The existence of repulsive force (in the case $\Delta>0$) allows the construction of more compact distribution when using anisotropic fluid than when using isotropic fluid.\par

Instead of solving eq. \eqref{2.1.13}, for any prescribed equation of state, we rather interested in solving eqs. \eqref{2.1.6}-\eqref{2.1.8} with the help of following ansatz \cite{Korkina81,Durgapal82},

\begin{equation}\label{2.1.14}
e^{\nu}=B_N\left(1+Cr^2\right)^N
\end{equation}\noindent
where $N$ is a positive integer and $B_N,\, C>0$ are two constants to be determined by the appropriate physical boundary conditions. Subtracting \eqref{2.1.6} from \eqref{2.1.7} one obtain the equation of ``pressure anisotropy",

\begin{eqnarray}\label{2.1.15}
\left(\frac{\nu^{''}}{2}-\frac{\nu^'\lambda^'}{4}+\frac{\nu^{'2}}{4}-\frac{\nu^'+\lambda^'}{2r}\right)e^{-\lambda}+\frac{\left(1-e^{-\lambda}\right)}{r^2}&&\nonumber\\
=\kappa(P_t-P_r)+\frac{2q^2}{r^4}&&
\end{eqnarray}\noindent
Equation \eqref{2.1.15} is a second order nonlinear differential equation in $\nu$ but first order linear in $\lambda$. At this moment it is convenient to introduce the following transformations

\begin{equation}\label{2.1.16}
e^{-\lambda}=Z,\,\, x=Cr^2
\end{equation}\noindent
which transform eqs. \eqref{2.1.6}-\eqref{2.1.8} to the following,

\begin{eqnarray}
\frac{\kappa}{C}P_r &=& \frac{[1+(2N+1)x]}{x(1+x)}Z-\frac{1}{x}+\frac{Cq^2}{x^2}\label{2.1.17}\\
\frac{\kappa}{C}P_t &=&\frac{(2N+N^2x)}{(1+x)^2}Z+\frac{[1+(1+N)x]}{(1+x)}\frac{dZ}{dx}-\frac{Cq^2}{x^2} \label{2.1.18}\\
\frac{\kappa}{C}\rho &=&-2\frac{dZ}{dx}-\frac{Z}{x}+\frac{1}{x}-\frac{Cq^2}{x^2} \label{2.1.19}
\end{eqnarray}\noindent
And the eq. \eqref{2.1.15} can be written in terms of auxiliary variable $x$ as,

\begin{eqnarray}\label{2.1.20}
\frac{dZ}{dx}&+&\left[\frac{(N^2-2N-1)x^2-2x-1}{x(1+x)(1+(1+N)x)}\right]Z\nonumber\\
&=&\frac{(1+x)}{x(1+(1+N)x)}\left(\frac{2Cq^2}{x}+\Delta x-1\right)
\end{eqnarray}\noindent
where, $\displaystyle\Delta=\frac{\kappa}{C}(P_t-P_r)$ is the measure of pressure anisotropy.
Eq. \eqref{2.1.20} yields the following solution \cite{Fatema-Murad13},

\begin{eqnarray}
Z&=&\frac{x}{(1+x)^{N-2}\left[1+(1+N)x\right]^{\frac{2}{1+N}}}\times\nonumber\\
&&
\int{\frac{(1+x)^{N-1}\left[1+(1+N)x\right]^{\frac{1-N}{1+N}}}{x^2}\left(\frac{2Cq^2}{x}+\Delta x\right)}\,dx\nonumber\\
&&+\frac{1}{(1+x)^{N-2}}-\frac{1}{2}\binom{N-1}{2}\frac{x}{(1+x)^{N-2}}\nonumber\\
&&-\sum_{i=0}^{N-4}\sum_{j=0}^{i+1}\frac{(-1)^j}{(N+1)^{i+2}}\binom{N-1}{i+3}\binom{i+1}{j}\times\nonumber\\
&&\frac{x[1+(N+1)x]^{i-j+1}}{(i-j+1)(1+x)^{n-2}}\nonumber\\
&&+A_N\frac{x}{(1+x)^{N-2}\left[1+(1+N)x\right]^{\frac{2}{1+N}}} \condition{$N\geq 4$}\label{2.1.21}
\end{eqnarray}\noindent
where $A_N$ is the constant of integration may be determined by imposing appropriate physical boundary conditions.

\numberwithin{equation}{subsection}
\subsection{Models of Electric Charge Distribution and Pressure Anisotropy}\label{Sec2.2}

As the ``realistic'' charge distribution inside the fluid sphere is not known \cite{Giuliani-Rothman08}, but it seems intuitively reasonable that due to electrical repulsion the charge distribution should be weighted toward the surface \cite{Anninos-Rothman01}. One can imagine several plausible mathematical forms of $2Cq^2/x^2$, to integrate the eq. \eqref{2.1.21}. Various authors presented variety of solutions previously for different suitable choices of charge distributions with isotropic pressure. Some of the solutions are compiled in table \ref{table1} (Also see \cite{Murad-Fatema13c}). In this work we consider the following forms of electric charge distribution and pressure anisotropy:
\begin{equation}\label{2.2.1}
\frac{2Cq^2}{x^2}=Kx^{n+1}(1+x)^{1-N}(1+mx)^p(1+(1+N)x)^{\frac{N-1}{N+1}}
\end{equation}
\begin{equation}
\Delta=\delta x(1-2ax)(1+x)^{1-N}(1+(1+N)x)^{\frac{N-1}{N+1}}
\end{equation}\label{2.2.2}\noindent
where  $K,\,\delta\geq0$, $n$ is a nonnegative integer, and $m,\,p$ are nonzero and $a$ is any real numbers. It must be emphasized that these hypothetical models of electric charge distribution and pressure anisotropy are chosen, in term of $x$, in such a way that these allow us to integrate the eq. \eqref{2.1.21} rather than for any particular physical reasons. Moreover, the electric field intensity and anisotropy vanish at the center and remains continuous and bounded in the interior of the star for a wide range of values of the parameters. Thus these choices may be physically reasonable and useful in the study of the gravitational behavior of anisotropic charged stellar objects.

\setlength{\heavyrulewidth}{1pt}
\setlength{\abovetopsep}{4pt}
\begin{table*}[h]
\centering
\caption{\label{table1} Exact static spherically symmetric perfect fluid solutions of Einstein-Maxwell equations obtained by different charge distributions for isotropic pressure $(\Delta=0)$. The second column shows the models which may be rediscovered from the present model \eqref{2.2.1}.}
\begin{tabular*}{\textwidth}{@{\extracolsep{\fill}}cccccll@{}}
\toprule
\multirow{2}{*}{$N$}&Generated by&\multirow{2}{*}{$n$}&\multirow{2}{*}{$m$}&\multirow{2}{*}{$p$}&
Charge Distribution&\multirow{2}{*}{Reference}\\
&present model&&&&$2Cq^2/x^2$&\\
\midrule
1&Y&0&$-$&0&$Kx$&\cite{Pant-Rajasekhara11}\\
1&Y&0&1&1&$Kx(1+x)$&\cite{Pant-Negi12}\\
1&Y&0&1&$n$&$Kx(1+x)^n$&\cite{Pant-Negi12}\\
2&Y&0&1&1&$Kx(1+3x)^{\frac{1}{3}}$&\cite{Pant-Rajasekhara11,Pant-Tewari11}\\
2&Y&0&1&2&$Kx(1+x)(1+3x)^{\frac{1}{3}}$&\cite{Pantetal.11a}\\
2&Y&0&1&3&$Kx(1+x)^2(1+3x)^{\frac{1}{3}}$&\cite{Pant-Faruqi12,Murad13}\\
2&Y&0&1&$n$&$Kx(1+x)^n(1+3x)^{\frac{1}{3}}$&\cite{Fatema-Murad13}\\
2&Y&$n$&$3$&$-1/3$&$K\displaystyle\frac{x^{n+1}}{(1+x)}$&\multirow{3}{*}{\cite{Murad-Fatema13c}}\\
2&N&$-$&$-$&$-$&$Kx^{n+1}(1+mx)^p(1+3x)^{\frac{1}{3}}$&\\
2&Y&$n$&$m$&$p$&$K\displaystyle\frac{x^{n+1}(1+mx)^p(1+3x)^{\frac{1}{3}}}{(1+x)}$&\\
2&N&$-$&$-$&$-$&$Kx^{n+1}(1+x)^{l-1}(1+3x)^{m+\frac{1}{3}}$&\cite{Murad-Fatema14}\\
2&N&$-$&$-$&$-$&$Kx(1+mx)^{\frac{1}{3}}(1+3x)^{\frac{1}{3}}$&\multirow{2}{*}{\cite{Rahman-Murad14}}\\
2&Y&0&$m$&$1/3$&$\displaystyle K\frac{x(1+mx)^{\frac{1}{3}}(1+3x)^{\frac{1}{3}}}{(1+x)}$&\\
3&Y&0&1&2&$Kx\sqrt{(1+4x)}$&\cite{Pantetal.11a}\\
3&Y&0&1&$n+2$&$Kx(1+x)^n\sqrt{(1+4x)}$&\cite{Pant-Maurya12}\\
4&Y&0&1&3&$Kx(1+5x)^{\frac{3}{5}}$&\cite{Pant11a}\\
4&Y&0&1&4&$Kx(1+x)(1+5x)^{\frac{3}{5}}$&\cite{Mehtaetal.13}\\
4&Y&0&1&$n+3$&$Kx(1+x)^n(1+5x)^{\frac{3}{5}}$&\cite{Murad-Fatema13a}\\
4&N&$-$&0&0&$\displaystyle\frac{Kx^r}{(1+x)^2}$&\cite{Mauryaetal.11}\\
4&Y&2&5&$-3/5$&$\displaystyle\frac{Kx^3}{(1+x)^3}$&\cite{Faruqi-Pant12}\\
5&Y&0&1&4&$Kx(1+6x)^{\frac{2}{3}}$&\cite{Gupta-Maurya11}\\
5&Y&0&1&5&$Kx(1+x)(1+6x)^{\frac{2}{3}}$&\cite{Fuloriaetal.11}\\
5&Y&0&1&6&$Kx(1+x)^2(1+6x)^{\frac{2}{3}}$&\cite{Fuloria-Tewari12}\\
5&Y&0&1&$n+3$&$Kx(1+x)^n(1+6x)^{\frac{2}{3}}$&\cite{Murad-Fatema13b}\\
6&Y&0&1&5&$Kx(1+7x)^{\frac{5}{7}}$&\cite{Maurya-Gupta11a}\\
$N$&Y&0&1&$N-1$&$n^2Kx[1+(N+1)x]^{\frac{(N-1)}{(N+1)}}$&\cite{Maurya-Gupta11c}\\
\bottomrule
\end{tabular*}
\end{table*}

\numberwithin{equation}{subsection}
\subsection{Anisotropic Charged Stellar Models}\label{Sec2.3}
Ishak et al. \cite{Ishaketal.01}, Lake \cite{Lake03}, and recently Maurya and Gupta \cite{Maurya-Gupta11b,Maurya-Gupta11c} showed that the ansatz for the metric function \eqref{2.1.14} produces an infinite family of analytic solutions of the self-bound type. Four of these were previously known ($N=$1, 3, 4, and 5). The most relevant case is for $N=2$, for which the speed of sound $\approx1/\sqrt{3}$ throughout most of the star, similar to the behavior of strange quark matter \cite{Lattimer-Prakash05}. For $N=2$ the solution of the Einstein-Maxwell system \eqref{2.1.6}-\eqref{2.1.8} for the model charge distribution and pressure anisotropy considered in this work are then given by,
\begin{description}

\item[]Case I: $p \neq -1$. $n=$ nonnegative integer
\begin{equation}\label{2.3.1}
e^{\nu}=B_2(1+x)^2
\end{equation}
\begin{eqnarray}\label{2.3.2}
  e^{-\lambda} &=& \frac{K}{m^{n+1}}\sum_{i=0}^n\frac{(-1)^i}{(n-i+p+1)}\binom{n}{i}\times\nonumber\\
&&\frac{x(1+mx)^{(n-i+p+1)}}{(1+3x)^{\frac{2}{3}}}+\delta \frac{x^2 (1-ax)}{(1+3x)^{\frac{2}{3}}}\nonumber\\
&&+1+A_2\frac{x}{(1+3x)^{\frac{2}{3}}} \end{eqnarray}
\begin{equation}\label{2.3.3}
\frac{2Cq^2}{x^2}=K\frac{x^{n+1}(1+mx)^p (1+3x)^{\frac{1}{3}}}{(1+x)}
\end{equation}
\begin{equation}\label{2.3.3}
\Delta=\frac{\delta x(1-2ax)(1+3x)^{\frac{1}{3}}}{(1+x)}
\end{equation}
\begin{eqnarray}\label{2.3.5}
  \frac{\kappa}{C}P_r &=& \frac{K}{m^{n+1}}\sum_{i=0}^n\frac{(-1)^i}{(n-i+p+1)}\binom{n}{i}\times\nonumber\\
&&\frac{(1+mx)^{(n-i+p+1)}(1+5x)}{(1+3x)^{\frac{2}{3}}(1+x)}\nonumber\\
&&+\frac{K}{2}\frac{x^{n+1}(1+mx)^p(1+3x)^{\frac{1}{3}}}{(1+x)}\nonumber\\
&&+\delta \frac{x(1-ax)(1+5x)}{(1+3x)^{\frac{2}{3}}(1+x)}\nonumber\\
&&+\frac{4}{(1+x)}+A_2\frac{(1+5x)}{(1+3x)^{\frac{2}{3}}(1+x)}
\end{eqnarray}
\begin{equation}\label{2.3.6}
 \frac{\kappa}{C}P_t= \frac{\kappa}{C}P_r+\Delta
\end{equation}
\begin{eqnarray} \label{2.3.7}
\frac{\kappa}{C}\rho &=& -\frac{K}{m^{n+1}}\sum_{i=0}^n\frac{(-1)^i}{(n-i+p+1)}\binom{n}{i}(1+mx)^{(n-i+p)}\times\nonumber\\
&&\frac{\Pi_{m,n,i,p}(x)}{(1+3x)^{\frac{5}{3}}}-\frac{K}{2}\frac{x^{n+1}(1+mx)^p (1+3x)^{\frac{1}{3}}}{(1+x)}\nonumber\\
&&-\delta x\frac{(5+(11-7a)x-17ax^2)}{(1+3x)^{\frac{5}{3}}}\nonumber\\
&&-A_2\frac{(3+5x)}{(1+3x)^{\frac{5}{3}}}
\end{eqnarray}

\item[]Case II: $p = -1$. $n = 0$
\begin{equation}\label{2.3.8}
e^{\nu}=B_2\left(1+x\right)^2
\end{equation}
\begin{eqnarray}\label{2.3.9}
e^{-\lambda}&=&\frac{K}{m}\frac{x\ln(1+x)}{(1+3x)^{\frac{2}{3}}}+\delta \frac{x^2 (1-ax)}{(1+3x)^{\frac{2}{3}}}+1\nonumber\\
&&+A_2\frac{x}{(1+3x)^{\frac{2}{3}}}
\end{eqnarray}
\begin{equation}\label{2.3.10}
\Delta=\frac{\delta x(1-2ax)(1+3x)^{\frac{1}{3}}}{(1+x)}
\end{equation}
\begin{equation}\label{2.3.11}
\frac{2Cq^2}{x^2}=K\frac{x(1+3x)^{\frac{1}{3}}}{(1+x)(1+mx)}
\end{equation}
\begin{eqnarray}\label{2.3.12}
\frac{\kappa}{C}P_r&=&\frac{K}{m}\frac{\ln(1+mx)(1+5x)}{(1+3x)^{\frac{1}{3}}(1+x)}
+\frac{K}{2}\frac{x(1+3x)^{\frac{1}{3}}}{(1+x)(1+mx)}\nonumber\\
&&+\delta\frac{x(1-ax)(1+5x)}{(1+3x)^{\frac{2}{3}}(1+x)}+\frac{4}{(1+x)}\nonumber\\
&&+A_2\frac{(1+5x)}{(1+3x)^{\frac{2}{3}}(1+x)}
\end{eqnarray}
\begin{equation}\label{2.3.13}
 \frac{\kappa}{C}P_t= \frac{\kappa}{C}P_r+\Delta
\end{equation}
\begin{eqnarray}\label{2.3.14}
\frac{\kappa}{C}\rho&=&-\frac{K}{m}\frac{(3+5x)\ln(1+mx)}{(1+3x)^{\frac{5}{3}}}
-\frac{2Kx}{(1+3x)^{\frac{2}{3}}(1+mx)}\nonumber\\
&&-\frac{K}{2}\frac{x(1+3x)^{\frac{1}{3}}}{(1+x)(1+mx)}\nonumber\\
&&-\delta x\frac{(5+(11-7a)x-17ax^2)}{(1+3x)^{\frac{5}{3}}}\nonumber\\
&&-A_2\frac{(3+5x)}{(1+3x)^{\frac{5}{3}}}
\end{eqnarray}\\

\item[]Case III: $p = -1$. $n=$ positive integer
\begin{equation}\label{2.3.15}
e^{\nu}=B_2\left(1+x\right)^2
\end{equation}
\begin{eqnarray}\label{2.3.16}
e^{-\lambda}&=&\frac{K}{m^{n+1}}\sum_{i=0}^{n-1}\left[\frac{(-1)^i}{(n-i)}\binom{n}{i}\frac{x(1+mx)^{(n-i)}}{(1+3x)^{\frac{2}{3}}}\right]\nonumber\\
&&+\frac{K}{m}(-1)^n\frac{x\ln(1+mx)}{(1+3x)^{\frac{1}{3}}}+\delta \frac{x^2 (1-ax)}{(1+3x)^{\frac{2}{3}}}\nonumber\\
&&+1+A_2\frac{x}{(1+3x)^{\frac{2}{3}}}
\end{eqnarray}
\begin{equation}\label{2.3.17}
\frac{2Cq^2}{x^2}=K\frac{x^{n+1}(1+3x)^{\frac{1}{3}}}{(1+x)(1+mx)}
\end{equation}
\begin{equation}\label{2.3.18}
\Delta=\frac{\delta x(1-2ax)(1+3x)^{\frac{1}{3}}}{(1+x)}
\end{equation}
\begin{eqnarray}\label{2.3.19}
\frac{\kappa}{C}P_r&=&\frac{K}{m^{n+1}}\sum_{i=0}^{n-1}\left[\frac{(-1)^i}{(n-i)}\binom{n}{i}\frac{(1+mx)^{(n-i)}(1+5x)}{(1+3x)^{\frac{2}{3}}(1+x)}\right]\nonumber\\
&&+\frac{(-1)^n K}{m^{n+1}}\frac{\ln(1+mx)(1+5x)}{(1+3x)^{\frac{2}{3}}(1+x)}\nonumber\\
&&+\frac{K}{2}\frac{x^{n+1}(1+3x)^{\frac{1}{3}}}{(1+x)(1+mx)}\nonumber\\
&&+\delta\frac{x(1-ax)(1+5x)}{(1+3x)^{\frac{2}{3}}(1+x)}\nonumber\\
&&+\frac{4}{(1+x)}+A_2\frac{(1+5x)}{(1+3x)^{\frac{2}{3}}(1+x)}
\end{eqnarray}
\begin{equation}\label{2.3.20}
 \frac{\kappa}{C}P_t= \frac{\kappa}{C}P_r+\Delta
\end{equation}
\begin{eqnarray}\label{2.3.21}
\frac{\kappa}{C}\rho&=&-\frac{K}{m^{n+1}}\sum_{i=0}^{n-1}\frac{(-1)^i}{(n-i)}\binom{n}{i}(1+mx)^{(n-i-1)}\times\nonumber\\
&&\frac{\Pi_{m,n,-1,i}(x)}{(1+3x)^{\frac{5}{3}}}-
\frac{(-1)^n K}{m^{n+1}}\frac{(3+5x)\ln(1+mx)}{(1+3x)^{\frac{5}{3}}}\nonumber\\
&&-\frac{(-1)^n K}{m^n}\frac{2x}{(1+mx)(1+3x)^{\frac{5}{3}}}\nonumber\\
&&-\frac{K}{2}\frac{x^{n+1}(1+3x)^{\frac{1}{3}}}{(1+mx)(1+x)}\nonumber\\
&&-\delta x\frac{(5+(11-7a)x-17ax^2)}{(1+3x)^{\frac{5}{3}}}\nonumber\\
&&-A_2\frac{(3+5x)}{(1+3x)^{\frac{5}{3}}}
\end{eqnarray}\noindent
where,
\begin{eqnarray*}
\Pi_{m,n,p,i}(x)&=&3+(2mn-2mi+2mp+5m+5)x\\\
&&+(6mn-6mi+6mp+11m)x^2.
\end{eqnarray*}
\end{description}
In the absence of electric field intensity $(K=0)$ and pressure anisotropy $(\delta=0)$
eqs. \eqref{2.3.1}-\eqref{2.3.21} reduce to the well-known Wyman solution\cite{Wyman49}, also known as Wyman IIa metric\footnote[2]{Wyman IIa metric is the generalization of Tolman VI solution. Leibovitz rediscovered Wyman IIa metric in \cite{Leibovitz69}. Adler \cite{Adler74}, Kuchowicz \cite{Kuchowicz75} and Adams-Cohen \cite{Adams-Cohen75} also rediscovered a particular case of Wyman IIa metric respectively, but none of those works cited Wyman's work!} according to the classification made in \cite{Delgaty-Lake98}. Hence, the models presented by the eqs. \eqref{2.3.1}-\eqref{2.3.21} represent the anisotropic charged analogues of Wyman-Adler-Kuchowicz solution. Equations \eqref{2.3.5}, \eqref{2.3.7}, \eqref{2.3.12}, \eqref{2.3.14}, and \eqref{2.3.19}, \eqref{2.3.21} constitute the equations of state of each case.

\numberwithin{equation}{section}
\section{Elementary Criteria for Physical Acceptability}\label{Sec3}

Due to the high nonlinearity of Einstein field equations \eqref{2.1.2} not many realistic physical solutions are known for the description of static spherically symmetric perfect fluid spheres. Out of 127 solutions only 16 were found to pass elementary tests of physical relevance \cite{Delgaty-Lake98}. A physically acceptable interior solution of the gravitational field equations must comply with the certain (not necessarily mutually independent) physical conditions \cite{Kuchowicz72,Buchdahl79}:

\begin{enumerate}
  \item[(a)] Regularity conditions
  \begin{itemize}
    \item[(i)]
     The solution should be free from physical and geometric singularities i.e. $e^{\nu}>0$ and $e^{\lambda}>0$ in the range $0\leq r\leq R$
    \item[(ii)]
     The radial and tangential pressures and density are positive, $P_r,\,P_t,\,\rho>0$.
    \item[(iii)]
     Radial pressure $P_r$ should be zero at boundary $r=R$ i.e. $P_r(r=R)=0$, the energy density and tangential pressure may follow $\rho(r=R)\geq0$ and $P_r(r=R)\geq0$.
  \end{itemize}
  \item[(b)] Stability conditions
  \begin{itemize}
    \item[(iv)]	
     In order to have an equilibrium configuration the matter must be stable against the collapse of local regions. This requires, \emph{Le Chatelier's principle} also known as local or microscopic stability condition, that the radial pressure $P_r$ must be a monotonically non-decreasing function of $\rho$ \cite{Bayin82},
                             \[dP_r/d\rho\geq0.\]
    \item[(v)]
     The relativistic adiabatic index is given by $\Gamma=\frac{(P_r+\rho)}{P_r}\frac{dP_r}{d\rho}$. The necessary condition for this exact solution to serve as a model of a relativistic star is that $\Gamma>\frac{4}{3}$.

  \end{itemize}
  \item[(c)] Causality condition
  \begin{itemize}
    \item[(vi)]
         The condition $0\leq\sqrt{dP_r/d\rho}\leq1$, $0\leq\sqrt{dP_t/d\rho}\leq1$ be the condition that the speed of sound not exceeds that of light.
\end{itemize}
  \item[(d)] Energy conditions
  \begin{itemize}
      \item[(vii)]
       A physically reasonable energy-momentum tensor has to obey the conditions $\rho\geq P_r+2P_t$ and $\rho+P_r+2P_t\geq0$.
  \end{itemize}
  \item[(e)] Monotone decrease of physical parameters
  \begin{itemize}
    \item[(ix)]
    	Pressure and density, should maximum at the center and monotonically decreasing towards the pressure free interface (i.e. boundary of the fluid sphere). Mathematically,
\begin{equation*}
\left(\frac{dP_r}{dr}\right)_{r=0}=0,\,\left(\frac{dP_t}{dr}\right)_{r=0}=0,\,\left(\frac{d\rho}{dr}\right)_{r=0}=0,
\end{equation*}
\begin{equation*}
 \left(\frac{d^2P_r}{dr^2}\right)_{r=0}<0,\left(\frac{d^2P_t}{dr^2}\right)_{r=0}<0,\, \left(\frac{d^2\rho}{dr^2}\right)_{r=0}<0
\end{equation*}
 So that, \[\frac{dP_r}{dr}<0,\,\frac{dP_t}{dr}<0,\,\frac{d\rho}{dr}<0, \, 0<r\leq R.\]
    \item[(x)]
    Additionally, the velocity of sound should be monotonically decreasing towards the surface, i.e., \[\frac{d}{dr}\left(\frac{dP_r}{d\rho}\right)<0,\,\frac{d}{dr}\left(\frac{dP_t}{d\rho}\right)<0\] for $0\leq r\leq R$.
    \item[(xi)]
    The ratio of pressure to density, $P_{r,t}/\rho$, should be monotonically decreasing with the increase of $r$, i.e., \[\frac{d}{dr}\left(\frac{P_r}{\rho}\right)_{r=0}=0,\,\,  \frac{d^2}{dr^2}\left(\frac{P_r}{\rho}\right)_{r=0}<0,\]
    \[\frac{d}{dr}\left(\frac{P_t}{\rho}\right)_{r=0}=0,\,\,  \frac{d^2}{dr^2}\left(\frac{P_t}{\rho}\right)_{r=0}<0.\]
  \end{itemize}
  \item[(f)] Matching condition
    \begin{itemize}
    \item[(xii)]
  	The interior solution should match continuously with an exterior Reissner-Nordstr\"{o}m solution,
\begin{eqnarray*}
  ds^2&=&\left(1-\frac{2M}{r}+\frac{Q^2}{r^2}\right)dt^2  -\left(1-\frac{2M}{r}+\frac{Q^2}{r^2}\right)^{-1}\times\\
&&dr^2-r^2\left(d\theta^2+\sin^2\theta d\varphi^2\right)\condition{$r\geq R.$}
\end{eqnarray*}
   This requires the continuity of $e^{\nu},\, e^{\lambda}$ and $q$ across the boundary $r=R$, \[e^{\nu(R)}=e^{{-\lambda(R)}}=\left(1-\frac{2M}{R}+\frac{Q^2}{R^2}\right)\] and $q(R)=Q$, where $M$ and $Q$ represent the total mass and charge inside the fluid sphere respectively.
  \end{itemize}
  \item[(g)] Charge distribution
  \begin{itemize}
    \item[(xiii)]
     Electric field intensity $E$, such that $E(0)=0$, is taken to be monotonically increasing, i.e., $dE/dr>0$ for $0<r\leq R$.
  \end{itemize}
  \item[(g)] Pressure anisotropy
  \begin{itemize}
    \item[(xiv)]
     Pressure anisotropy, $\Delta$, vanishes at the center, i.e., $\Delta(0)=0$ \cite{Bowers-Liang74,Ivanov02a}.
  \end{itemize}
  \item[(h)] Allowable mass to radius ratio
    \begin{itemize}
    \item[(xv)]
    	Buchdahl \cite{Buchdahl59} obtained an absolute constraint of the maximally allowable mass-to-radius ratio $(M/R)$ for isotropic fluid spheres of the form $2M/R\leq8/9$ (in the unit, $c = G = 1$) which states that, for a given radius a static isotropic fluid sphere cannot be arbitrarily massive. B\"{o}hmer and Harko \cite{Bohmer-Harko07} proved that for a compact object with charge, $Q(<M)$, there is a lower bound for the mass-radius ratio,
       \[\frac{3Q^2}{2R^2}\frac{\left(1+\frac{Q^2}{18R^2}\right)}{\left(1+\frac{Q^2}{12R^2}\right)}\leq \frac{2M}{R}\]
    Upper bound of the mass of charged sphere was generalized by Andr\'{e}asson \cite{Andreasson09} and proved that \[\sqrt{M}\leq\frac{\sqrt{R}}{3}+\sqrt{\frac{R}{9}+\frac{Q^2}{3R}}\]

  \end{itemize}

\end{enumerate}

\numberwithin{equation}{section}
\section{Physical Boundary Conditions}\label{Sec4}
\subsection{Determination of the Arbitrary Constant $A_2$}\label{Sec4.1}
The boundary condition $P_r(R)=0$, can be utilized to specify $A_2$. For the \emph{Case I}:
\begin{dmath*}
A_2=-\frac{K}{m^{n+1}}\sum_{i=0}^n\left[\frac{(-1)^i}{(n-i+p+1)}\binom{n}{i}(1+mX)^{(n-i+p+1)}\right]
-\frac{K}{2}\frac{X^{n+1}(1+mX)^p (1+3X)}{(1+5X)}-\delta X(1-aX)-4\frac{(1+3X)^{\frac{2}{3}}}{(1+5X)} \end{dmath*}\noindent
where $X=CR^2$.

\numberwithin{equation}{subsection}
\subsection{Total Charge to Radius Ratio $Q/R$}\label{Sec4.2}
Using $X=CR^2$ in eq. \eqref{2.2.1} we obtain the square of ratio $Q/R$,
\begin{dmath}\label{4.2.1}
\frac{Q^2}{R^2}=\frac{K}{2}\frac{X^{n+2} (1+mX)^p (1+3X)^{\frac{1}{3}}}{(1+X)}
\end{dmath}

\numberwithin{equation}{subsection}
\subsection{Total Mass to Radius Ratio $M/R$}\label{Sec4.3}
By matching the metric coefficients obtained in \eqref{2.3.1}-\eqref{2.3.2} with the exterior Reissner-Nordstr\"{o}m metric at the boundary and with reference to the eq. \eqref{4.2.1} one can establish the equation of compactness,
\begin{equation}\label{4.3.1}
\frac{2M}{R}=\left(1-e^{-\lambda(X)}+\frac{Q^2}{R^2}\right)
\end{equation}

\numberwithin{equation}{subsection}
\subsection{Total Charge to Mass Ratio $Q/M$}\label{Sec4.4}

By the use of the equations \eqref{4.2.1} and \eqref{4.3.1} we obtain the charge to mass ratio $Q/M$,

\numberwithin{equation}{subsection}
\subsection{Determination of the Constant $B_2$}\label{Sec4.5}

The constant $B_2$ can be specified by the boundary condition $e^{\nu(R)}=e^{-\lambda(R)}$, which gives,
\begin{equation}\label{4.5.1}
B_2=(1+X)^{-2}e^{-\lambda(X)}
\end{equation}

\numberwithin{equation}{subsection}
\subsection{Central and Surface Redshifts}\label{Sec4.6}
The central and surface redshifts of the charged fluid sphere are given by
\[z_c=\sqrt{e^{-\nu(0)}}-1, \,\, z_s=\sqrt{e^{-\nu(R)}}-1=\frac{(1+X)^{-1}}{\sqrt{B_2}}-1\]

\numberwithin{equation}{section}
\section{Construction of Physically Realistic Fluid Spheres}\label{Sec5}

\numberwithin{equation}{subsection}
\subsection{Pressure and Density Gradients}\label{subsec5.1}
Differentiating the pressure and density equations \eqref{2.3.5}-\eqref{2.3.7}, \eqref{2.3.12}-\eqref{2.3.14}, \eqref{2.3.19}-\eqref{2.3.21} with respect to the auxiliary variable $x$ one obtains the pressure and density gradients respectively for each of the model EOS.

\begin{description}
 \item[]\label{Case I} Case I: $p\neq-1,\,n=0$
  \begin{eqnarray}
    \frac{\kappa}{C}\frac{dP_r}{dx}&=&\frac{K}{m(p+1)}(1+mx)^p\frac{\Sigma(x)}{(1+3x)^{\frac{5}{3}}(1+x)^2}\nonumber\\
 &&+\frac{K}{2}(1+mx)^{p-1}\frac{\Phi(x)}{(1+3x)^{\frac{5}{3}}(1+x)^2}\nonumber\\
 &&+\delta\frac{\Psi_1(x)}{(1+x)^2(1+3x)^{\frac{5}{3}}}-\frac{4}{(1+x)^2}\nonumber\\
 &&+2A_2\frac{(1-5x^2 )}{(1+3x)^{\frac{5}{3}}(1+x)^2}\\ \label{5.1.1}
\frac{\kappa}{C}\frac{dP_t}{dx}&=&\frac{\kappa}{C}\frac{dP_r}{dx}+\frac{d\Delta}{dx}\\ \label{5.1.2}
\frac{\kappa}{C}\frac{d\rho}{dx}&=&-\frac{K}{m(p+1)}(1+mx)^{p-1}\frac{\Theta(x)}{(1+3x)^{\frac{8}{3}}}\nonumber\\
&&+\frac{K}{2}(1+mx)^{p-1}\frac{\Phi(x)}{(1+3x)^{\frac{5}{3}}(1+x)^2}\nonumber\\
&&-\delta\frac{\Psi_2(x)}{(1+3x)^{\frac{8}{3}}}+10A_2\frac{(1+x)}{(1+3x)^{\frac{8}{3}}}\label{5.1.3}
  \end{eqnarray}
where,
\begin{eqnarray*}
\Sigma(x)&=&(mp+m-1)+(9mp+8m-20)x\\
&&+(23mp+3m-35)x^2+(15mp-20m)x^3, \\
\Phi(x)&=&1+(mp+m+4)x+(4mp+4m+1) x^2\\
&&+(3mp+m) x^3,\\
\Theta(x)&=&(5mp+5m-10)+(2m^2p^2+7m^2p+5m^2\\
&&+22mp+2m-10)x+(12m^2p^2+34m^2p\\
&&+12m^2+21mp+m)x^2+(18m^2p^2+39m^2p\\
&&+11m^2)x^3,\\
\Psi_1(x)&=&1+(11-2a)x+(23-20a)x^2\\
&&+(5-46a)x^3-25ax^4\\
\Psi_2(x)&=&5+(12-14a)x+(11-58a)x^2-68ax^3\\
\frac{d\Delta}{dx}&=&\delta\frac{(1+(4-4a)x+(1-16a)x^2-8ax^3)}{(1+x)^2(1+3x)^{\frac{2}{3}}}
\end{eqnarray*}
\end{description}

\numberwithin{equation}{subsection}
\subsection{Specifying the Maximum Mass and Radius}\label{Sec5.2}
A fluid sphere satisfying conditions (a) and (e) of Sect. \ref{Sec3} will be termed as \emph{well-behaved}. For a particular set $(m,n,p,a,\delta)$ the values of $K,\,X$ have been plugged-in to the eqs. \eqref{2.3.5}-\eqref{2.3.7} and \eqref{5.1.1}-\eqref{5.1.3} for which the fluid distribution satisfies the elementary criteria for physical acceptability. Once the compactness $M/R$ and the ratio $Q/R$ of the compact fluid sphere are obtained the maximum mass can then be calculated by using one of the following quantities: (i) radius, (ii) central density, (iii) surface density, (iv) central pressure or, (v) total charge as parameter. In this subsection we describe how to calculate the values various physical variables. For the \emph{Case I} this can be accomplished in the following way:

\numberwithin{equation}{subsubsection}
\subsubsection{For a given radius}\label{Sec5.2.1}
    \begin{itemize}
    \item[(a)] Total Mass
    \begin{equation}\label{5.2.1.1}
M=\frac{R}{2}\left(1-e^{-\lambda(X)}+\frac{Q^2}{R^2}\right)
    \end{equation}
where the mass $M$ is in the unit km \footnote[3]{The following physical constants, in their conventional values, have been used for the numerical calculation: $c=1=2.997\times10^8$m s$^{-1}$, $G=1=6.674\times10^{-11}$N m$^2$ kg$^{-2}$, $M_\odot=1.486$ km$\,=2\times10^{30}$kg}.
    \item[(b)] Total Charge
        \begin{equation}\label{5.2.1.2}
        Q=R\sqrt{\frac{K}{2}\frac{X^2(1+mX)^p(1+3X)^{\frac{1}{3}}}{(1+X)}}
        \end{equation}
where the charge $Q$ is in the unit of radius\footnote[4]{$Q$ km$=\left(Q\times1000\times c^2/\sqrt{\frac{G}{4\pi\epsilon_0}}\right)\,C$}.
    \item[(c)] Central Density
        \begin{equation}\label{5.2.1.3}
        \rho_c=\frac{3X}{\kappa R^2}\left[\frac{-K}{m(p+1)}-A_2\right]
        \end{equation}

    \item[(d)] Surface Density
        \begin{equation}\label{5.2.1.4}
        \rho_s=\frac{X}{\kappa R^2}\Omega(X)
        \end{equation}\noindent
        where,
        \begin{eqnarray*}
        \Omega(X)&=&-\frac{K}{m(p+1)}(1+mX)^p\frac{\Pi_{m,0,p,0}(X)}{(1+3X)^{\frac{5}{3}}}\\
&&-\frac{K}{2}\frac{X(1+mX)^p(1+3X)^{\frac{1}{3}}}{(1+X)}\\
&&-\delta X\frac{(5+(11-7a)X-17aX^2)}{(1+3X)^{\frac{5}{3}}}-A_2\frac{(3+5X)}{(1+3X)^{\frac{5}{3}}}
        \end{eqnarray*}
    \end{itemize}

\numberwithin{equation}{subsubsection}
\subsubsection{For a given surface density}\label{Sec5.2.2}
        \begin{equation}\label{5.2.2.1}
        R^2=\frac{X}{\kappa\rho_s}\Omega(X)
        \end{equation}
    And the total mass, total charge, and the central density can be calculated by eqs. \eqref{5.2.1.1}-\eqref{5.2.1.3}.

\numberwithin{equation}{subsubsection}
\subsubsection{For a given central density}\label{Sec5.2.3}
    The radius of the charged fluid sphere for a prescribed central density can be calculated by the following equation,
        \begin{equation}\label{5.2.3.1}
        R^2=\frac{3X}{\kappa\rho_c}\left[\frac{-K}{m(p+1)}-A_2\right]
        \end{equation}
        where the central energy density $\rho_c$ is given in the unit kg m$^{-3}$  and the radius in m. The total mass, total charge, and the surface density then can be calculated by eqs. \eqref{5.2.1.1}, \eqref{5.2.1.2}, and \eqref{5.2.1.4} respectively.

\numberwithin{equation}{subsubsection}
\subsubsection{For a given central pressure}\label{Sec5.2.4}
    The radius of the charged fluid sphere for a prescribed central pressure can be calculated by the following equation.
       \begin{equation}\label{5.2.4.1}
       R^2=\frac{X}{\kappa P_c}\left[\frac{K}{m(p+1)}+4+A_2\right]
       \end{equation}\noindent
where the central pressure $P_c$ is given in the unit N m$^{-2}$ \footnote[5]{1N m$^{-2}=10$ dyne cm$^{-2}$ and $1$ MeV fm$^{-3}=1.6022\times10^{33}$ dyne cm$^{-2}$}. The total mass, total charge, the central and surface densities then can be calculated by eqs. \eqref{5.2.1.1}-\eqref{5.2.1.4}.

\numberwithin{equation}{subsubsection}
\subsubsection{For a given electric charge}\label{Sec5.2.5}
    The radius of the charged fluid sphere for a prescribed total charge can be calculated by the following equation
       \begin{equation}\label{Sec5.2.5.1}
        R=Q/\sqrt{\frac{K}{2}\frac{X^2(1+mX)^p (1+3X)^{\frac{1}{3}}}{(1+X)}}
       \end{equation}
    where the charge $Q$ is given in the unit km. Then the total mass, total charge, the central and surface densities then can be calculated by eqs. \eqref{5.2.1.1}-\eqref{5.2.1.4}.

\numberwithin{equation}{subsection}
\subsection{Physical Analysis of the Models}\label{Sec5.3}
For each choice of constant parameters $(K,\,m,\,n,\,p,\,\delta,\,a)$,  the maximum mass of charged star depends on the corresponding set of maximum values of $X=X_{\mathrm{max}}$ upto which the pressure and density and their gradients satisfy $P_r\geq0$, $P_t\geq0$, $\rho\geq0$, $dP_r/dx<0$, $dP_t/dx<0$, $d\rho/dx<0$ and the speed of sound satisfy $0\leq\sqrt{dP_r/d\rho}\leq1$, $0\leq\sqrt{dP_t/d\rho}\leq1$ and monotonically decreasing with increasing radius.\par

\begin{description}
  \item[Case Ia:] \emph{Isotropic pressure}\\
We set $(m,\,n,\,p,\,\delta,\,a)=(10^4,\,0,\,0.24,\,0,\,0)$. For this choice the range of values, $K\geq0.093,\,0<X\leq0.672$ are obtained over which the fluid distribution satisfies the above mentioned inequalities. With the decrease of $K$, $X$ increases. The maximum value of compactness parameter is obtained $(2M/R)_{\mathrm{max}}=0.8246$, using \eqref{4.3.1}, at $K_{\mathrm{min}}=0.093,\,X_{\mathrm{max}}=0.672$. Corresponding to the values of $(K_{\mathrm{min}},\,X_{\mathrm{max}})$ the total charge to radius ratio, and total charge to total mass ratio are found to be $Q/R=0.3878$ and $Q/M=0.9406$ using eq. \eqref{4.2.1}. For a particular choice of stellar surface density $\rho_s = 4.68\times10^{14}$ g cm$^{-3}$ \footnote[6]{The surface density of bare strange stars is equal to that of strange quark matter (SQM) at zero pressure. By using the formula given in \cite{Zdunik00} the SQM density with $m_s c^2=150$ MeV, $\alpha_c=0.17$, $B = 60$ MeV fm$^{-3}$ is calculated to be $\rho_s=4.6\times10^{14}$ g cm$^{-3}$. It is therefore some fourteen orders of magnitude larger than the surface density of normal neutron stars.} as parameter the total mass and other physical quantities are calculated by the use of eqs. \eqref{5.2.1.1}-\eqref{5.2.1.3} and found $M_{\mathrm{max}}=2.8740M_\odot$, $R=10.35$ km, $P_c=279.73$ MeV fm$^{-3}$, $\rho_c=2.52\times10^{15}$ g cm$^{-3}$, $Q=4.66\times10^{20}\,C$.\\

\item[Case Ib:] \emph{Anisotropic pressure}\\
Due to the presence of pressure anisotropy the range of values of $K$ and $X$ are increased. For the input parameters $m=10^4$, $n=0$, $p=0.23$, $\delta=0.2$, $a=0.71$ the maximum value of compactness parameter is obtained $(2M/R)_{\mathrm{max}}=0.8307$, at $K_{\mathrm{min}}=0.094,\,X_{\mathrm{max}}=0.7$. The charge-radius ratio, and charge-mass ratio are found to be $Q/R=0.3890$ and $Q/M=0.9366$. For $\rho_s = 4.68\times10^{14}$ g cm$^{-3}$ the maximum mass found $M_{\mathrm{max}}=2.8999M_\odot$, with radius $R=10.37$ km, central pressure $P_c=284.30$ MeV fm$^{-3}$, central energy density $\rho_c=2.65\times10^{15}$ g cm$^{-3}$, and total charge $Q=4.68\times10^{20}\,C$. Details are reported in Tables \ref{table2}-\ref{table3}.

\end{description}

\setlength{\heavyrulewidth}{1.5pt}
\setlength{\abovetopsep}{4pt}
\begin{table*}[tb]
\small
\caption{\label{table2}Some values of parameters $(K,X_{\mathrm{max}})$ for which well-behaved charge fluid sphere can be generated}
\begin{tabular*}{\textwidth}{@{\extracolsep{\fill}}ccccccccccccc@{}}
\toprule
$p$&$m$&$\delta$&$a$&$\left(K_{\mathrm{min}},X_{\mathrm{max}}\right)$&$A_2$&$B_2$&$\left(P_r/c^2\rho\right)_c$&$\sqrt{\left(dP_r/c^2d\rho\right)_c}$ &$\sqrt{\left(dP_t/c^2d\rho\right)_c}$&$2M/R$&$Q/R$&$Q/M$\\
\midrule
\multicolumn{13}{c}{\emph{Case I} $n=0$}\\
\midrule
0.20&$10^4$&0&0&(0.093,\,0.672)&-2.5122&0.1165&0.1974&0.5887&0.5887&0.8246&0.3878&0.9406\\
0.22&$10^4$&0.2&0.71&(0.103,\,0.699)&-2.5488&0.1111&0.1898&0.5684&0.5619&0.8306&0.3891&0.9368\\
0.23&$10^4$&0.2&0.71&(0.094,\,0.700)&-2.5439&0.1109&0.1908&0.5696&0.5631&0.8307&0.3891&0.9366\\
0.24&$10^4$&0.2&0.71&(0.086,\,0.700)&-2.5403&0.1109&0.1915&0.5706&0.5641&0.8306&0.3889&0.9365\\
\midrule
\multicolumn{13}{c}{\emph{Case II} $n=0$}\\
\midrule
\multirow{2}{*}{$-1$}&$2$&0&0&(2.063,\,0.612)&-2.9967&0.1295&0.1116&0.4086&0.4086&0.8042&0.3751&0.9327\\
                     &$2$&0.2&0.64&(1.866,\,0.708)&-2.9714&0.0988&0.1154&0.4116&0.4056&0.8638&0.3899&0.9028\\
\bottomrule
\end{tabular*}
\end{table*}

\setlength{\heavyrulewidth}{1.5pt}
\setlength{\abovetopsep}{4pt}
\begin{table*}[t]
\centering
\small
\caption{\label{table3}Maximum mass and the various physical variables of charged fluid spheres for a given surface density}
\begin{threeparttable}
\begin{tabular*}{\textwidth}{@{\extracolsep{\fill}}cccccccccc@{}}
\toprule
\multirow{2}{*}{$p$}&\multirow{2}{*}{$m$}&\multirow{2}{*}{$\delta$}&\multirow{2}{*}{$a$}&\multirow{2}{*}{$\left(K,X_{\mathrm{max}}\right)$} &\multicolumn{5}{c}{$\rho_{s,14}$\tnote{a}  $=4.68$}\\
\cline{6-10}
&&&&&$M(M_\odot)$&$R$(km)&$P_{c,35}$\tnote{b}&$\rho_{c,15}$\tnote{c}&$Q_{20}$\tnote{d}\\
\midrule
\multicolumn{10}{c}{\emph{Case I} $n=0$}\\
\midrule
0.24&$10^4$&0&0&$(0.093,\,0.672)$&2.8741&10.35&4.48&2.52&4.66\\
0.22&$10^4$&0.2&0.71&$(0.103,\,0.699)$&2.8988&10.37&4.53&2.66&4.68\\
0.23&$10^4$&0.2&0.71&$(0.094,\,0.700)$&2.8999&10.37&4.55&2.66&4.68\\
0.24&$10^4$&0.2&0.71&$(0.086,\,0.700)$&2.8998&10.37&4.56&2.65&4.68\\
\midrule
\multicolumn{10}{c}{\emph{Case II} $n=0$}\\
\midrule
\multirow{2}{*}{$-1$} &2&0&0&(2.063, 0.612)&2.7739&10.25&2.81&2.80&4.46\\
                      &2&0.2&0.64&(1.866, 0.708)&2.8613&10.14&3.40&3.28&4.59\\
\bottomrule
\end{tabular*}
     \begin{tablenotes}
       \item[a] $\rho_s=\rho_{s,14}\times10^{14}$g cm$^{-3}$
       \item[b] $P_c=(P_r)_c=(P_t)_c=P_{c,35}\times10^{35}$dyne cm$^{-2}$
       \item[c] $\rho_c=\rho_{c,15}\times10^{15}$g cm$^{-3}$
       \item[d] $Q=Q_{20}\times10^{20}\,C$
     \end{tablenotes}
\end{threeparttable}
\end{table*}

\begin{figure}[b]
\includegraphics[width=\columnwidth]{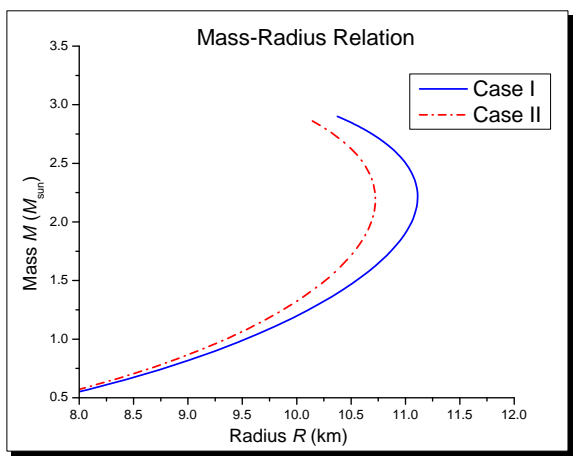}
\caption{\label{Figure1} Mass-radius relation for a sequence of charged fluid spheres. The solid (blue) line corresponds to the \emph{Case I} with the input $(m,n,p,\delta,a,K_{\mathrm{min}})=$ $(10^4,0,0.23,0.2,0.71,0.094)$, $0<X\leq0.7$. And the dashed-doted (red) line corresponds to the \emph{Case II} $(2,\,0,\,-1,\,0.2,\,0.64,\,1.866)$, $0<X\leq0.708$ with surface density $\rho_s=4.69\times10^{14}$ g cm$^{-3}$.}
\end{figure}

\begin{figure}
\includegraphics[width=\columnwidth]{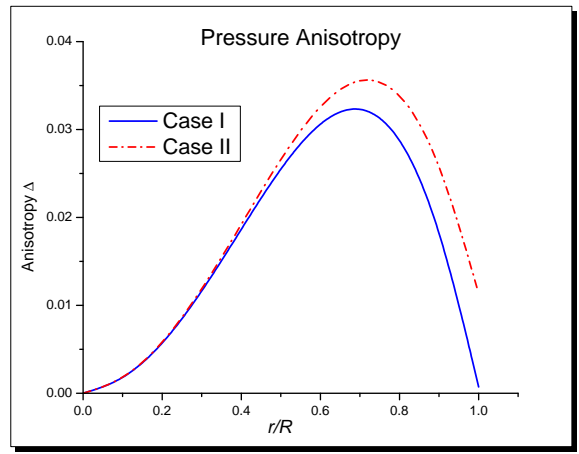}
\caption{\label{Figure2} Behaviour of pressure anisotropy $\Delta$ for the same sequences of charged fluid spheres in fig. \ref{Figure1}.}
\end{figure}

\begin{figure}
\includegraphics[width=\columnwidth]{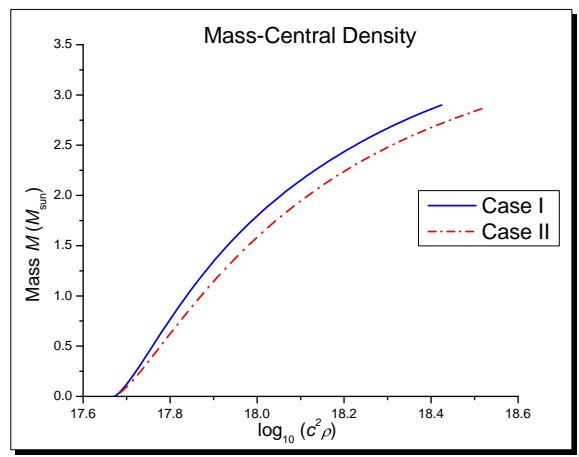}
\caption{\label{Figure3} Mass vs. central density for same sequences of charged fluid spheres in fig. \ref{Figure1}.}
\end{figure}

The behaviors of various physical variables in the interior of the star have been investigated and found regular and well behaved throughout the fluid sphere. It has been observed that the radial and tangential speeds of sound $\sqrt{dP_r/d\rho}$, $\sqrt{dP_t/d\rho}$ always remain less than the speed of light and the condition of causality is satisfied. The mass-radius relation for a sequence of charged fluid spheres generated by the \emph{Case I} with the input $m=10^4$, $n=0$,
$p=0.2$, $\delta=0.2$, $a=0.71$, $K_{\mathrm{min}}=0.094$, $0<X\leq0.7$ and \emph{Case II} with input $m=2$, $n=0$, $p=-1$, $\delta=0.2$, $a=0.64$, $K=1.866$ $0<X\leq0.708$ with surface density $\rho_s=4.69\times10^{14}$ g cm$^{-3}$ have been demonstrated in Fig. \ref{Figure1}. The behavior of Fig. \ref{Figure1} reproduces that of other quark star models \cite{Negreirosetal.09}. The behaviors of pressure anisotropy the these cases have been demonstrated in fig \ref{Figure2}. In Fig. \ref{Figure3} the mass versus central density for the same sequences of configurations are plotted and this shows the necessary condition of stability is satisfied $(dP_r/d\rho_c>0)$. \par
\begin{figure}
\includegraphics[width=\columnwidth]{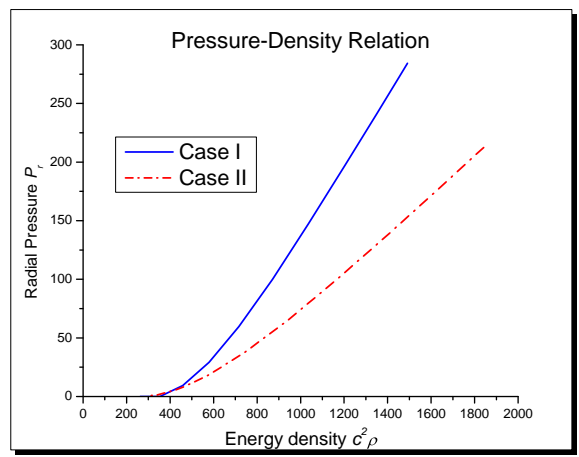}
\caption{\label{Figure4} Pressure-density profiles for the same sequences of charged fluid spheres in fig. \ref{Figure1}.}
\end{figure}
\begin{figure}
\includegraphics[width=\columnwidth]{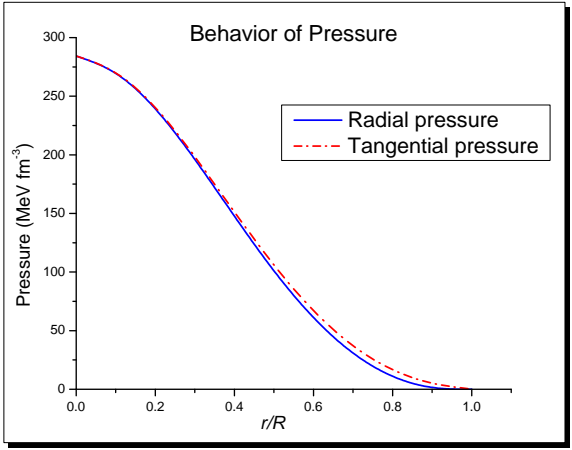}
\caption{\label{Figure5} Behavior of radial pressure $P_r$ and tangential pressure $P_t$ (MeV fm$^{-3}$) for the \emph{Case I} with the input $(m,n,p,\delta,a,K_{\mathrm{min}},X_{\mathrm{max}})=$ $(10^4,0,0.23,0.2,0.71,0.094,0.7)$. The solid (blue) line corresponds to radial pressure and the dashed-doted (red) line corresponds to the tangential pressure.}
\end{figure}
\begin{figure}
\includegraphics[width=\columnwidth]{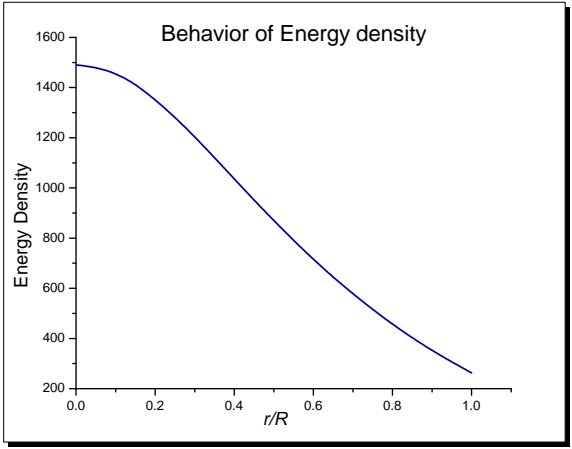}
\caption{\label{Figure6} Behavior of energy density $c^2 \rho$ (MeV fm$^{-3}$) for the same charged fluid sphere as in fig. \ref{Figure5}.}
\end{figure}
\begin{figure}
\includegraphics[width=\columnwidth]{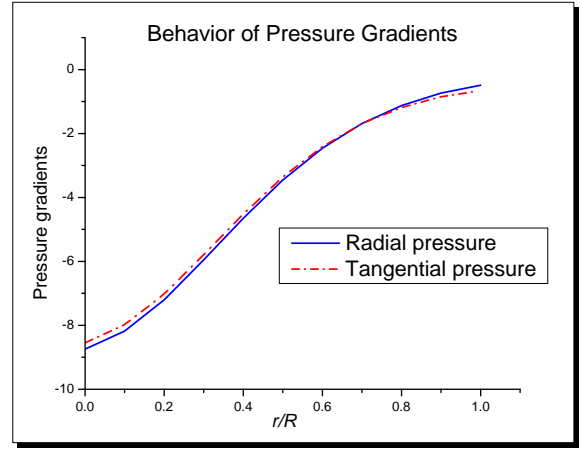}
\caption{\label{Figure7} Behavior of radial and tangential pressure gradients $\kappa dP_r/Cdx,\,\kappa dP_t/Cdx$ for the same charged fluid sphere as in fig. \ref{Figure5}.}
\end{figure}
\begin{figure}
\includegraphics[width=\columnwidth]{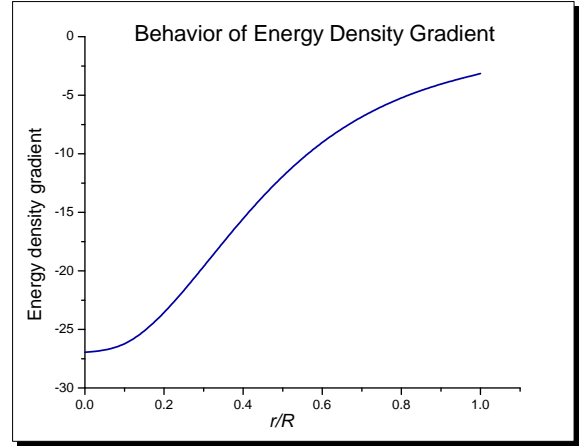}
\caption{\label{Figure8} Behavior of energy density gradient $\kappa d\rho/Cc^2dx$ for the same charged fluid sphere as in fig. \ref{Figure5}.}
\end{figure}
\begin{figure}
\includegraphics[width=\columnwidth]{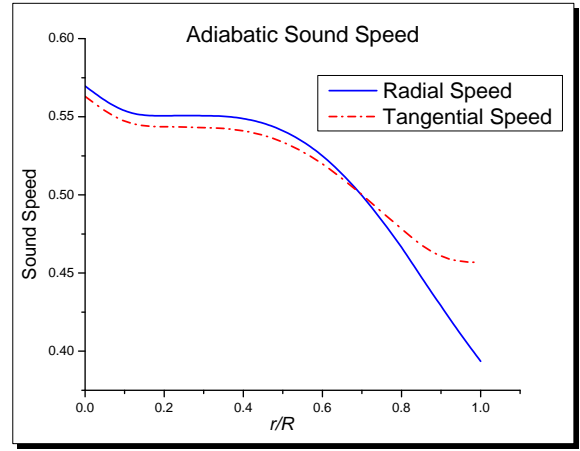}
\caption{\label{Figure9} Speed of adiabatic sound $\sqrt{dP_r/c^2d\rho},\, \sqrt{dP_t/c^2d\rho}$ for the strange star candidates in fig. \ref{Figure5}.}
\end{figure}

The pressure-density profiles together with various other physical variables given by our analytical model, are plotted in Fig. \ref{Figure4}-\ref{Figure9}. We have demonstrated that the models obtained in subsection \ref{Sec2.3} satisfy the physical requirements for wide range of values of $m,\,n,\,p,\,\delta,\,a$ and $K$, giving us a possibility for different charge variations and anisotropy within the fluid spheres. The resulting spheres can be utilized to construct physically reasonable compact self-bound charged stellar model such as charged strange quark star.

\numberwithin{equation}{section}
\section{An Application of the Model for Some Well Known Strange Star Candidates}\label{Sec6}

The analysis of very compact astrophysical objects has been a key issue in relativistic astrophysics for the last few decades. Recent observations show that the estimated mass and radius of several compact objects such as X-ray pulsar Her X-1, X-ray burster 4U 1820-30, millisecond pulsar SAX J 1808.4-3658, X-ray sources 4U 1728-34, PSR 0943+10 and RX J185635-3754 are not compatible with the standard neutron star models \cite{Deyetal.98,Lietal.99}. For a recent review on this may readers are referred to \cite{Weber05}.\par
Based on the analytic model developed so far, to get an estimate of the range of various physical parameters of some potential strange star candidates we have calculated the values of the relevant physical quantities, such as central/surface pressure and density, by using the refined mass and predicted radius of 12 pulsars recently reported in \cite{Gangopadhyayetal.13}. The values are reported in Table \ref{table4}.\par

\setlength{\heavyrulewidth}{1.5pt}
\setlength{\abovetopsep}{4pt}
\begin{table*}[tb]
\centering
\small
\caption{\label{table4}Physical values of energy density and pressure for different strange stars calculated by \emph{Case I} with $m=10^4,\,n=0$.}
\begin{tabular*}{\textwidth}{@{\extracolsep{\fill}}llcccccc@{}}
\toprule
Strange  &  &  &  &  &  & &\\
Star & $(p,\delta,a,K,X)$ & $M(M_\odot)$ & $R$ (km) & $P_{c,35}$ & $\rho_{c,15}$ & $\rho_{s,14}$ & $Q_{20}$\\
Candidate  &  &  &  &  &  & &\\
\midrule
\multirow{2}{*}{PSR J1614-2230}	& (0.24, 0, 0, 0.093, 0.396)	& \multirow{2}{*}{1.97}	 &
\multirow{2}{*}{9.69}	& 2.88  & 1.74 & 6.20 & 2.50\\
          	& (0.23, 0.2, 0.71, 0.094, 0.403)	&  & 	& 2.87  & 1.79	& 6.15 &2.45\\
\multirow{2}{*}{PSR J1903+327}	& (0.24, 0, 0, 0.093,0.304)	& \multirow{2}{*}{1.667}	 &
\multirow{2}{*}{9.438}	& 2.17  & 1.46 & 6.44 & 1.83\\
          	& (0.23, 0.2, 0.71, 0.094, 0.307)	&  & 	& 2.14  & 1.49	& 6.39 &1.78\\
\multirow{2}{*}{Vela X-1}	& (0.24, 0, 0, 0.093, 0.333)	& \multirow{2}{*}{1.77}	 &
\multirow{2}{*}{9.56}	& 2.38  & 1.54 & 6.33 & 2.05\\
          	& (0.23, 0.2, 0.71, 0.094, 0.337)	&  & 	& 2.36  & 1.58	& 6.29 &2.00\\
\multirow{2}{*}{4U 1820-30}	  & (0.24, 0, 0, 0.093, 0.295)	& \multirow{2}{*}{1.58}	 &
\multirow{2}{*}{9.1}	& 2.25  & 1.54 & 6.89 &1.71\\
          	& (0.23, 0.2, 0.71, 0.094, 0.298)	&  & 	& 2.22  & 1.57	& 6.85 &1.67\\
\multirow{2}{*}{4U 1608-52}	  & (0.24, 0, 0, 0.093, 0.324)	& \multirow{2}{*}{1.74}  &
\multirow{2}{*}{9.528}	& 2.32  & 1.52 & 6.36 &1.98\\
          	& (0.23, 0.2, 0.71, 0.094, 0.328)&  & 	& 2.29  & 1.55	& 6.32 &1.94\\
\bottomrule
\end{tabular*}
\end{table*}

\DeclareGraphicsExtensions{.pdf}

\section{Concluding Remarks}\label{Sec7}

In this work we have studied particular simple families of relativistic charged stellar models obtained by solving Einstein-Maxwell field equations for a static spherically symmetric distribution of perfect fluid distribution based on three \emph{ad hoc} assumptions, one for metric potential and others for the forms of electric charge distribution and pressure anisotropy. The analytical equation of state has been computed from the resulting metric. These families of analytical relativistic stellar models may be considered as \emph{anisotropic charged analogues} of Wyman-Adler solution. \par
A wide range of values of constant parameters are allowed to specify the maximum mass of charged fluid spheres. Various authors usually have chosen $2\times10^{14}$g cm$^{-3}$ as stellar surface density to calculate the mass and radius of the charged fluid spheres which may have given rise to the stellar configuration as massive as $4-6M_\odot$ with much lower central density. Such massive configuration may not serve as a realistic model for a strange quark star. This choice is, therefore, not a physical one. Modeling a compact (quark) star requires the use of a higher surface density. Certainly, the value of the surface density affects the calculated value of the stellar mass - to see this, we observe that the method employed in the present work one can obtain arbitrarily large maximum mass just by inserting vanishing small surface density (e.g., $0.1-1$ g cm$^{-3}$ to model a thin crust). In our model calculation, the density at the stellar radius is chosen within the range $4-10\times10^{14}$g cm$^{-3}$ \cite{Bombaci01} and drops abruptly to zero, as with all stellar models matching an interior metric to the external Reissner-Nordstr\"{o}m form.  This sharp drop in density is a reasonable model approximation since the thickness of the ``quark surface" is of order $1$ fm, a negligibly small dimension compared to the stellar radius.\par
In this work we assumed $P_t>P_r\,(\Delta>0)$ and have shown that the upper bound on the maximum mass increases in the presence of anisotropy. Moreover, for the some models (Case I), the speed of sound is obtained $\approx 1/\sqrt{3}$ at the center and remains almost the same throughout most of the fluid sphere. This behavior is like MIT bag model. An analytical stellar model with such physical features is most likely to present realistic model of strange quark stars. And hence the EOS given by our models, besides the usual linear EOS based on phenomenological MIT bag model, could play a significant role in the description of internal structure of electrically charged \emph{bare} strange quark stars.

\begin{acknowledgements}
One of the authors (M.H. Murad) expresses his sincere gratitude to Prof. Mofiz Uddin Ahmed, and Prof. A A Ziauddin Ahmad, Chair, Department of Mathematics and Natural Sciences, BRAC University, Dhaka, Bangladesh for their motivations and encouragements.
\end{acknowledgements}

\bibliographystyle{spphys}        
\bibliography{bibliography}       
\end{document}